\documentclass[aps,prl,twocolumn,groupedaddress,showpacs,floatfix]{revtex4}
\usepackage{dcolumn}
\usepackage{amsmath}
\usepackage{amssymb}
\usepackage{graphicx}
\usepackage{bm}
\usepackage[T1]{fontenc}
\usepackage{color}
\usepackage{url}
\usepackage[bf]{subfigure}
\usepackage{rotating}

\begin{document}

\title{Seeking for Simplicity in Complex Networks, and Its Consequences
for Cascade Failures}

\author{Luciano da Fontoura Costa}
\email{luciano@if.sc.usp.br}
\author{Francisco A. Rodrigues}
\affiliation{Instituto de F\'{\i}sica de S\~{a}o Carlos, Universidade de S\~{a}o Paulo, Av. Trabalhador
S\~{a}o Carlense 400, Caixa Postal 369, CEP 13560-970, S\~{a}o Carlos,
S\~ao Paulo, Brazil}

\begin{abstract}
Complex networks can be understood as graphs whose connectivity
deviates from those of regular or near-regular graphs, which are
understood as being `simple'.  While a great deal of the attention so
far dedicated to complex networks has been duly driven by the
`complex' nature of these structures, in this work we address the
identification of simplicity, in the sense of regularity, in complex
networks.  The basic idea is to seek for subgraphs exhibiting small
dispersion (e.g. standard deviation or entropy) of local measurements
such as the node degree and clustering coefficient.  This approach
paves the way for the identification of subgraphs (patches) with
nearly uniform connectivity, therefore complementing the
characterization of the complexity of networks. We also performed
analysis of cascade failures, revealing that the removal of vertices
in `simple' regions results in smaller damage to the network structure
than the removal of vertices in the heterogeneous regions. We
illustrate the potential of the proposed methodology with respect to
four theoretical models as well as protein-protein interaction
networks of three different species. Our results suggest that the
simplicity of protein interaction grows as the result of natural
selection. This increase in simplicity makes these networks more
robust to cascade failures.
\end{abstract}


\maketitle

The rise of the complex networks research area was ultimately
motivated by the finding that graphs obtained from natural or
human-made structures tended to present intricate connectivity when
compared to regular (e.g. lattices and meshes) or nearly-regular
graphs (e.g.  Flory and Erd\H{o}s-R\'enyi networks). Given their
structured connectivity, complex networks have provided excellent
models for several `complex' real-world systems ranging from the
Internet to protein-protein interaction (e.g.~\cite{Boccaletti05,
Costa:survey}).  Yet, the dicotomy between simplicity and complexity
continues to provide the motivation for related investigations.

The purpose of the present work is precisely to develop means to
identify simplicity and regularity in complex networks in order not
only to obtain better characterization of such structures, but also to
investigate how such features can affect respective dynamics such as
cascade failures.  More specifically, the motivation for finding
regularities in complex networks includes but is not limited to the
following issues: (1) the properties and distribution of regular
patches can help the characterization, understanding and modeling of
complex networks; (2) the presence of regular patches in a complex
network may suggest a hybrid nature of that network (e.g. a
combination of the regular and scale-free paradigms); and (3) regular
patches can have strong influence on dynamical processes taking place
on the network.

\begin{figure}[htb]
  \begin{center}
  \subfigure[]{\includegraphics[width=0.22\linewidth]{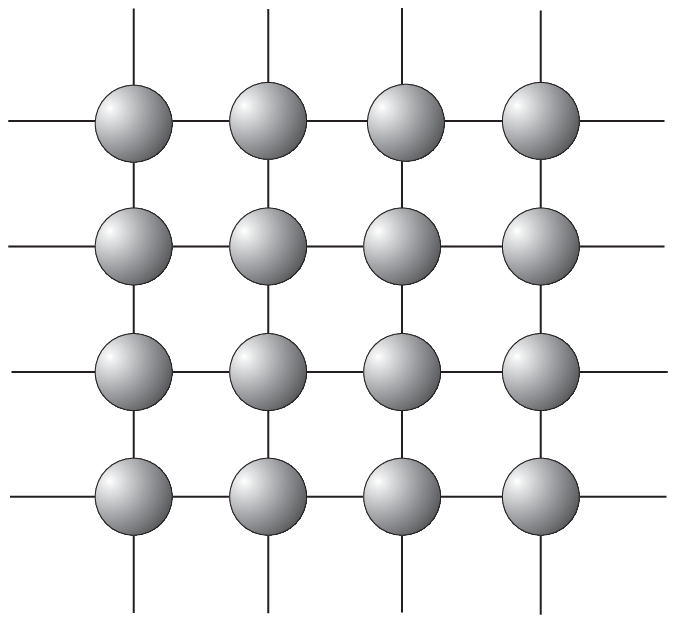}}
  \subfigure[]{\includegraphics[width=0.2\linewidth]{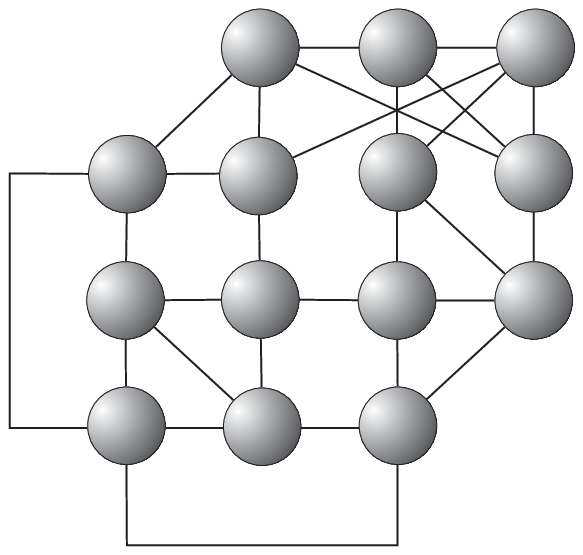}}
  \subfigure[]{\includegraphics[width=0.32\linewidth]{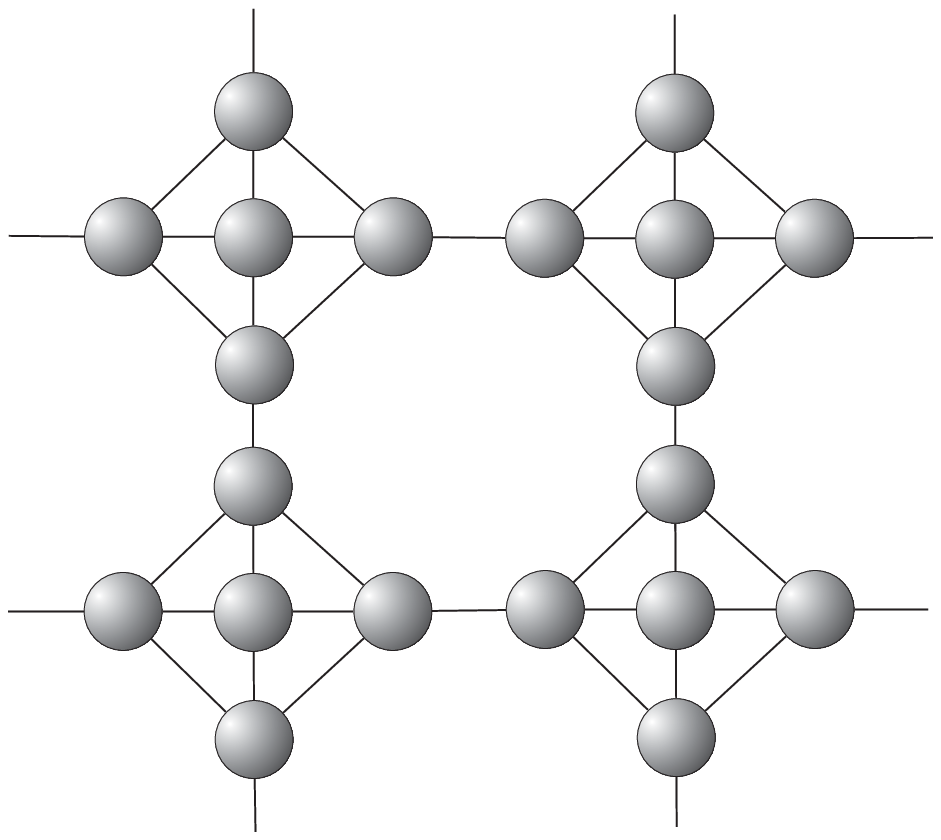}}
  \caption{Three examples of `regular' graphs, with all nodes with degree
       equal to 4.
  ~\label{fig:samedegree}}
  \end{center}
\end{figure}

In graph theory, a \emph{regular graph} is characterized by having all
its nodes with exactly the same degree.  However, such a definition is
limited in the sense that it is still possible to obtain a large
diversity of such `regular' graphs.  Figure~\ref{fig:samedegree}
presents three examples of `regular' graphs. All nodes in these graphs
exhibit the same degree, equal to four.  While the structures in (a)
and (c) have more `regular' connectivity, the graph in (b) looks
rather irregular.  Actually, even the structure in (c) has
irregularities if we consider measurements such as the individual
clustering coefficient.  It is clear from such an example that
identical node degree is not enough to ensure uniform connectivity.

As we are interested in characterizing and identifying regularity
(i.e. `simplicity') in complex networks, the first important step
consists of stating clearly what is meant by regularity, uniformity or
simplicity.  Perhaps the most strict imposition of regularity on a
network is that every node would be undistinguishable from any other
node whatever the considered measurements.  Therefore, identical
measurements (e.g. node degree, clustering coefficient, hierarchical
degrees, etc.) would be obtained for any of the nodes in the perfectly
regular network.  In this sense, regularity becomes closely related to
symmetries in the networks.

However, as we want to achieve increased flexibility, in this work we
relax the above definition and propose that a graph (or subgraph) is
\emph{regular} whenever all its nodes present similar values for a set
of measurements.  Therefore, this definition is relative to the
allowed degree of measurements dispersion, as well as the selection of
the own set of measurements.  Interestingly, traditional regularity in
graph theory can be understood as a particular instance of the above
definition in the case of null dispersion of node degree.  Here we
allow some tolerance for the variability of the measurements (e.g. in
order to cope with incompleteness or noise during the network
construction).

In order to characterize the local vertex structure, we considered the
degree, clustering coefficient, neighboring degree and the locality
index. The \emph{degree} of each of any of its nodes $i$, henceforth
abbreviated as $k(i)$, is defined as the number of immediate neighbors
of $i$. The \emph{clustering coefficient} of that node, represented as
$cc(i)$, is calculated by dividing the number of edges between the
immediate neighbors of $i$, i.e. $n_E(i)$, by the maximum possible
number of connections between those nodes, i.e.  $n_E(i) =
k(i)(k(i)-1)/2$. The \emph{average neighboring degree} $r(i)$
(e.g.~\cite{Maslov_Sneppen:2002}) of a node $i$, corresponds to the
average of the degrees of the immediate neighbors of $i$. The
\emph{locality index} $\mathrm{loc}(i)$ of a
node $i$ corresponds to the ratio between the number of connections
among the set of nodes comprising $i$ and its immediate neighbors
divided by the sum of the edges connected to all those
nodes~\cite{outliers:2006}. This measurement has been motivated by the
\emph{matching index}~\cite{Kaiser_Hilgetag:2004}, which is adapted
here in order to reflect the `internality' of the connections of all
the immediate neighbors of a given reference node, instead of a single
edge.

In order to find the regular patches, the set of measurements defining
homogeneity needs to be previously selected. Though such a set may
include just the node degree (compatible with the traditional concept
of regular graphs), because we want to impose more strict demands on
regularity it is necessary to consider additional features describing
the network connectivity around each node. Each vertex $i$ is
represented by a vector with $M$ measurements, $X_i$, which is
normalized~\cite{Costa:survey} in order to have zero means and unit
standard deviation. Then, the network, which is represented by the set
of such vectors $X = \{X_1,X_2,\dots,X_N\}$ is projected into a
two-dimensional space by considering principal component analysis --
PCA (e.g.~\cite{Duda_Hart, Johnson_Wichern:2002, Costa_Cesar,
Costa:survey}). Such a statistical mapping implements a linear
transformation (actually a rotation in the phase space) that ensures
that the maximum dispersion of the points will be achieved along the
initial projection axes (i.e. those corresponding to the largest
absolute values of eigenvalues of the covariance matrix of the data).
In addition, such a transformation optimally removes the redundancy of
the data, which is fundamental in our analysis since local
measurements are known to be correlated in most real-world
networks~\cite{Ravasz03:PRE}.

Having chosen the set of measurements and defined the projection, it
is necessary to identify those nodes which are characterized by small
dispersion of the measurements. Vertices with similar topological
features tend to be mapped close one another in the two-dimensional
space. However, visual inspection can provide inaccurate results
because of the form of the distribution of points in the
projection. Thus, we estimated the probability density in the 2D space
by considering the non-parametric Parzen windows
approach~\cite{parzen1962epd,Duda_Hart}. This method involves
convolving the feature vectors (represented as Dirac's deltas in the
2D projected space) with a two-variated Gaussian function (a normal
distribution), allowing the interpolation of the probability
density. In this way, high concentrations of points yield peaks in the
probability density, which correspond to respective classes of
vertices with homogeneous connectivity.

After determining the probability density of the node measurements as
mapped into the two-dimensional projection, it is necessary to
identify the obtained clusters of regular nodes.  This is performed
starting with the highest peak of the density.  A cluster is created
and associated to this peak, and its value is assigned to a control
variable $V$.  The value of $V$ is successively decreased and used to
threshold the density, from which eventual new peaks are searched.  In
case a new peak appears, a new cluster is defined. Whenever two peaks
merge, as a consequence of the progressive reduction of $V$, their
respective clusters are subsumed, creating a branch in the
hierarchical structure of clusters.  When $V$ reaches its final value
of 0, a tree representing the progressive merging of the peaks (and
respective clusters) is obtained.  The more significative clusters are
identified by taking the clusters corresponding to the longest
segments in the obtained tree.

However, the nodes defining a cluster do not necessarily correspond to
a regular patch, as they might not be connected in the original
graph. Therefore, the last step in the regular patch detection
corresponds to obtaining the connected subgraphs for each considered
homogeneous class in the distribution.  Three indicators of the
regularity of the network under analysis are considered in the current
work: (i) the number of detected peaks $P$, (ii) the relative size of
the maximum component identified considering all detected peaks, and
(iii) index of dispersion of the measurements inside the largest
regular region. The value of $P$ defines the number of different
structures that can be found in the network. These structure can be
thought of as generalized types of motifs, because they do not have
regular pre-defined structures~\cite{Milo:2002}, but are statistically
similar. 
In order to compare results obtained for networks with
different sizes, we define the \emph{simplicity coefficient} as the
the ration between the number of vertices in the largest connected region ($S$)
and the network size ($N$),
\begin{equation}
\Sigma = \frac{S}{N}.
\end{equation}
In addition, since the regularity can vary inside the largest
regular region, we can define a super-regularity coefficient. The
elements of the first eigenvector associated to the largest
eigenvalue, $\vec{v}_1$, provide the dispersion of each respective
measurement. Therefore, the level of regularity of the region can be
quantified in terms of the coefficient of variation of the elements
of such eigenvector, \emph{i.e.} the \emph{super-regularity
coefficient} can be given by
\begin{equation}
\Gamma = \frac{\sigma_{\vec{v_1}}}{\langle \vec{v_1} \rangle} \left(
\frac{1}{1 + \frac{\sigma_{\vec{v_1}}}{\langle \vec{v_1} \rangle},}
\right)
\end{equation}
where $\langle \vec{v_1} \rangle$ is the average and
$\sigma_{\vec{v_1}}$ is the standard deviation of $\vec{v_1}$. It
can be shown that $0\leq \Gamma \leq 1$. Note that networks
presenting values of $\Sigma$ and $\Gamma$ close to one tend to be
highly regular regarding all the considered measurements.

\begin{figure}[ht]
\begin{center}
\subfigure[]{\includegraphics[width=0.6\linewidth]{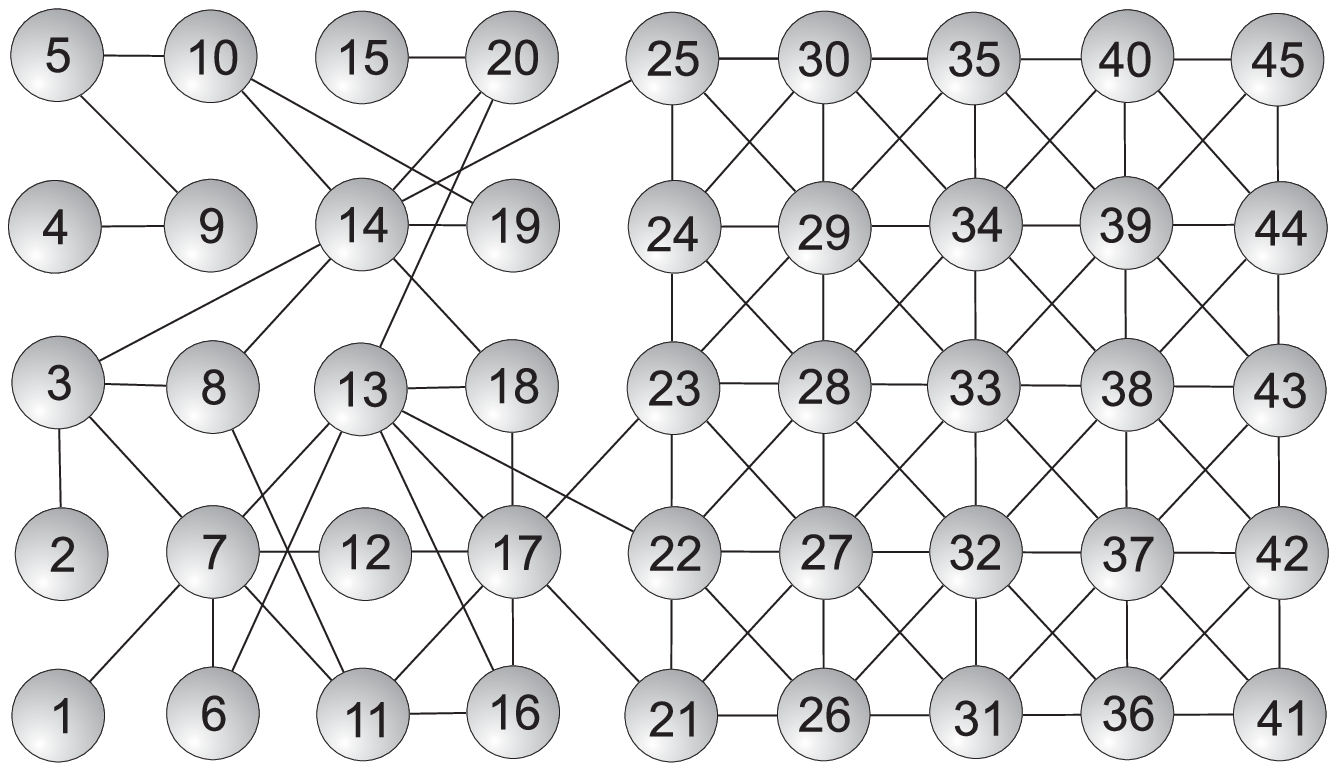}}\\
\subfigure[]{\includegraphics[width=0.45\linewidth]{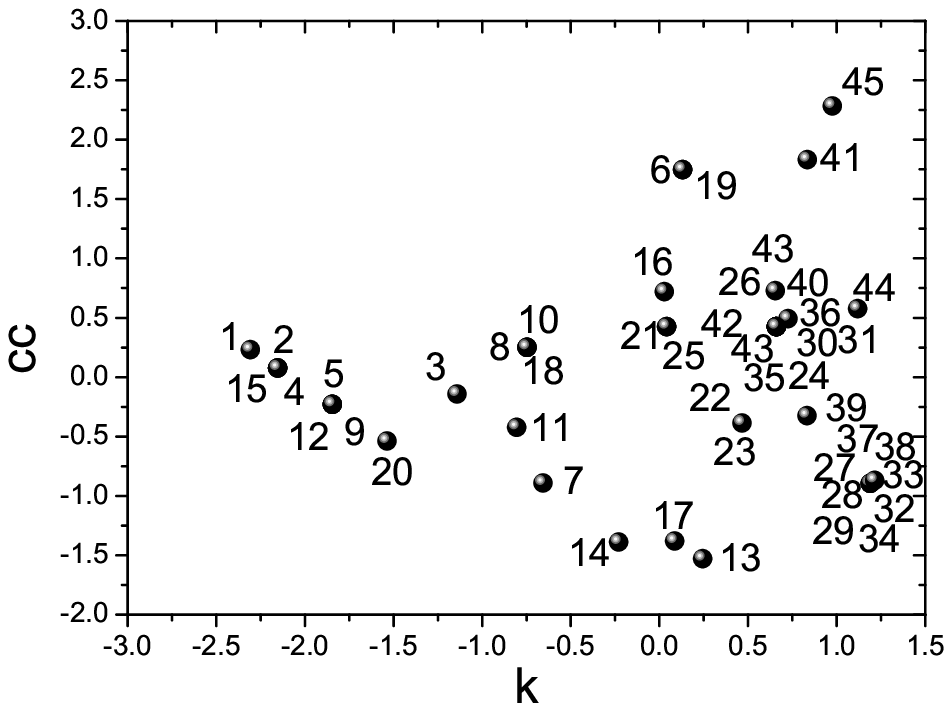}}\hspace{0.2cm}
\subfigure[]{\includegraphics[width=0.5\linewidth]{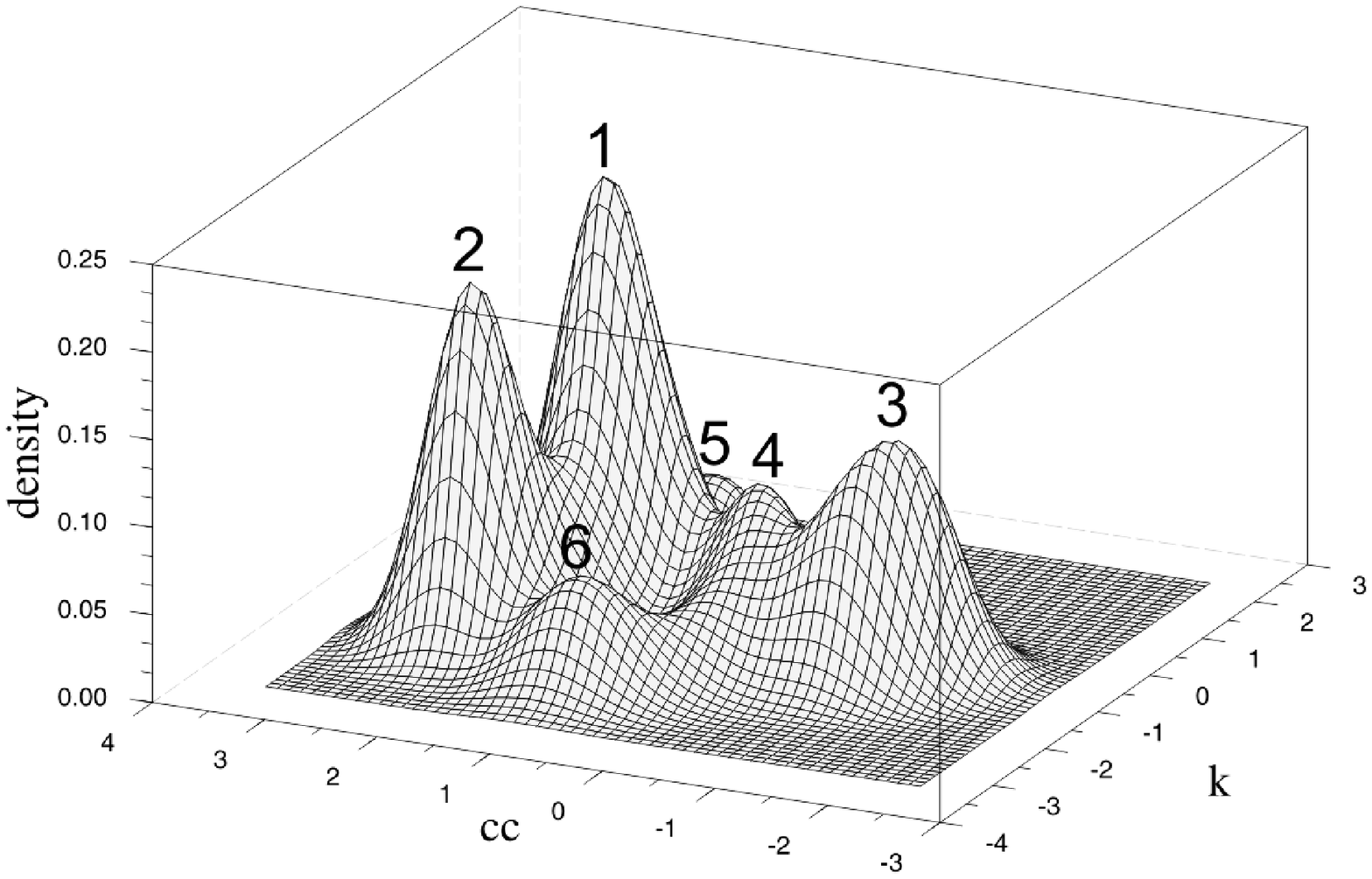}}\\
\subfigure[]{\includegraphics[width=0.45\linewidth]{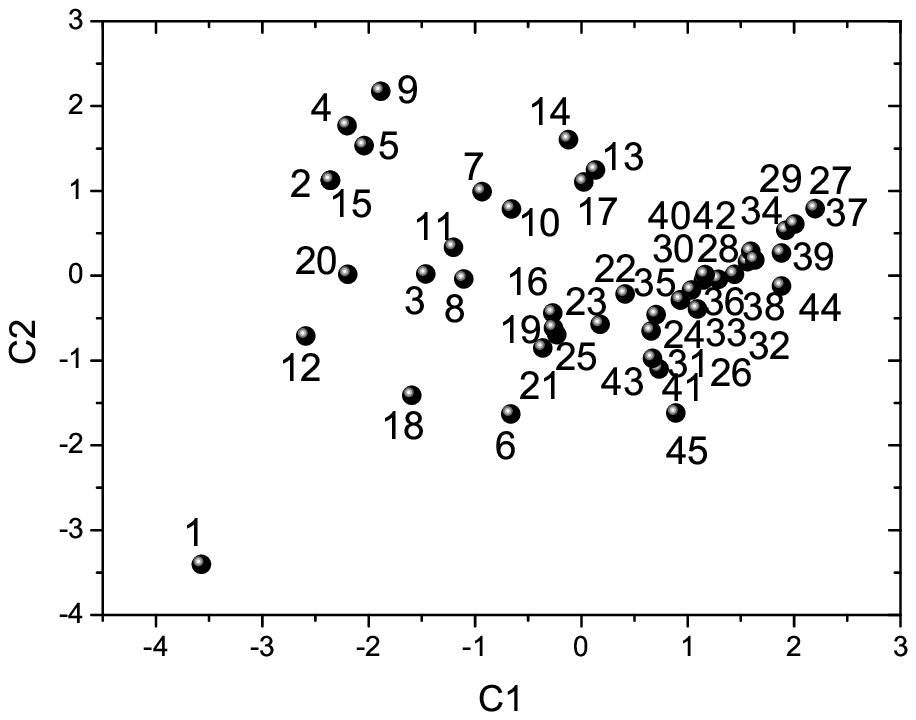}}\hspace{0.2cm}
\subfigure[]{\includegraphics[width=0.5\linewidth]{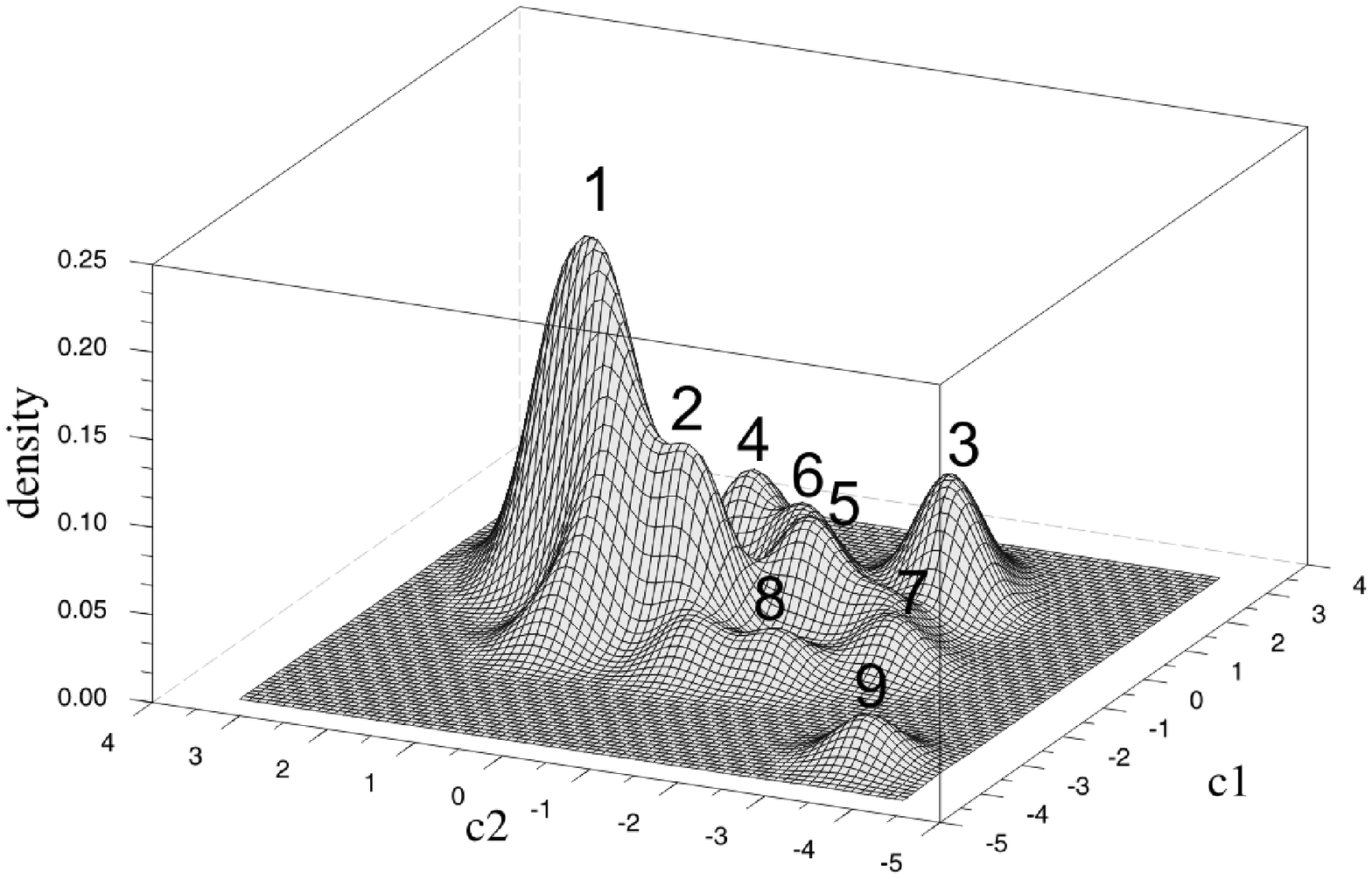}}
\caption{Illustration of the method to determine the simple patches
in the network(a). (b) Distribution of the
degree and cluster coefficient of each vertex and (c) the respective
probability density function. The PCA projection of the measurements
$M = \{k(i),cc(i),r(i), loc(i)\}$ into the two-dimensional space (d)
and the respective probability density (e).  The numbers in (c) and
(e) indicate the order of the peaks, from the highest to the
lowest.} \label{Fig:example}
\end{center}
\end{figure}

We illustrate the above methodology by considering the network in
Figure~\ref{Fig:example}(a). First, a set of measurements is extracted
and the vertices are projected into a two dimensional space. Note that
the projection using principal component analysis is necessary only
when more than two measurements are considered. We took into
account two different configurations of measurements: (i) $M_1 =
\{k(i),cc(i) \}$, and (ii) $M_2 = \{k(i),cc(i),r(i),loc(i)\}$. In
the first case, the two main regular regions are composed by the
vertices $R1 = \{21,24,25,26,30,31,35,36,40,42,43,44\}$ and $R2 =
\{22, 23, 27, 28, 29, 32, 33, 34, 37, 38, 39\}$. While most of the
vertices corresponding to region $R1$ belong to the border of the
regular region of Figure~\ref{Fig:example}(a), the vertices of $R2$
are internal to that region. In the latter case, i.e.\ considering a
larger set of measurements, the connected main region corresponds to
the vertices $R = \{22, 23, 24, 26, 27, 28, 29, 30, 31, 32, 33, 34,
35, 36, 37, 38, 39, \\40, 41, 42, 43, 44, 45\}$, which forms the
regular region presented in Figure~\ref{Fig:example}(a).  Our analysis
of network models and real-world networks took into account these
four local measurements.

In order to verify the importance of simple regions in networks under
dynamic aspects, we considered the analysis of cascade failures.
Cascade failures result in avalanche of breakdowns over the network
when nodes and links are sensitive to overloading~\cite{Motter02:PRE,
Motter04:PRL}. For a given network, a quantity of information (or
energy) can be interchanged between pairs of nodes following the
shortest paths distances at each time step. The capacity of a node
$i$, $C_i$, is proportional to its initial load $L_i$, $C_i = (1+
\alpha)L_i$, which is the maximum load that $i$ can handle. We
represent the load $i$ by the betweenness
centrality~\cite{Motter02:PRE, Motter04:PRL}. When a single node is
removed from the network, the dynamics of redistribution of flows
starts over the network and cascades can be triggered. Indeed, such
removals change the shortest paths between nodes and, consequently,
the distribution of the loads, creating overloads on some nodes. For
$\alpha=0$, it is guaranteed that at time $t=0$ no node is overloaded
and the system is working properly.  Larger values of $\alpha$
increase the capacity of nodes and reduce the chance of cascade
breakdown.

The analysis of cascade fails is performed by monitoring the
avalanches when a single node in the regular and irregular regions
is removed, independently. The damage caused by a cascade is
quantified in terms of the relative size $G$ of the largest
component, $G = N_f/N$, where $N_f$ is the size of the largest
component after the avalanche.

\begin{figure}[t]
\begin{center}
\includegraphics[width=0.85\linewidth]{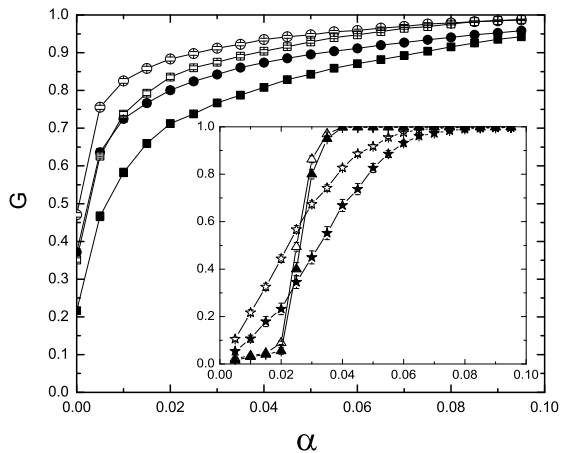} \\
\caption{Cascading failure in BA (circles) and NL (squares) network models,
and in ER (stars) and KT (triangles) (in the inset).
The removal of vertices in the simple regions are represented by
white symbols and in the heterogeneous regions by black symbols. Each
curve corresponds to an average of $G$ after removing every vertex in
each region, while the errors bars represent the standard deviations. }
\label{Fig:cascade}
\end{center}
\end{figure}

Figure~\ref{Fig:cascade} shows the value of $G$ when a single vertex
is removed in the largest simple and non-simple (the remainder of the
network) regions for the Erd\H{o}s-R\'{e}nyi (ER), Knitted
(KT)~\cite{Costa2007:condmat}, Barab\'{a}si-Albert (BA) and
Krapivsky-Redner (NL) (based on non-linear preferential attachment,
$P_{i\rightarrow j} = k_j^{0.5}/\sum_u
k_u^{0.5}$)~\cite{Krapivsky01:PRE} theoretical network models. Each
point in the scatterplot corresponds to an average of the relative
size of the resulting component $G$ after removing of each vertex $i$
in the simple and non-simple regions.

As we can see, the removal of vertices in the
simple region tends to cause smaller damage on the network than the
removal of vertices in the heterogeneous region for all models. This
property had already been observed considering the whole network, in
which homogenous networks tend to be more robust under fails and
target attacks~\cite{Motter02:PRE, Motter04:PRL}.  Therefore, the
simpler regions are fundamental to network robustness. In addition,
the curves obtained for models with similar structures, as BA e NL,
present similar behaviors. The same happens with the ER and KT network
models --- the latter corresponding to the most regular model in our
analysis~\cite{Costa2007:condmat}, which is reflected in
Table~\ref{table1}.

The protein databases were obtained from the Biogrid repository for
protein interactions~\cite{stark2006bgr}. The analysis of the
protein-protein interaction networks of the progressively more evolved
species \emph{Sacharomyces cerevisiae}, \emph{Drosophila melanogaster}
and \emph{Homo sapiens} allowed us to investigate how the simplicity
of the connections has changed during evolution under natural
selection. As we can see in Table~\ref{table1}, the level of
simplicity clearly increases with evolution --- the protein
interaction network of the \emph{H. sapiens} is the most regular. The
super-regularity coefficient also increases with the complexity of the
organism. A possible explanation of this remarkable phenomena is
related to protein evolution. Since hubs tend to evolve more slowly
than less connected proteins, because they are more important in the
organism, the addition of new connections due to mutation and
duplication tend to favor the proteins that are not hubs, therefore
increasing regularity~\cite{fraser2002erp}. This process implies that
more robust networks are obtained for more complex
organisms. Considering the cascade effect on the protein interaction
networks, we observed that the removal of vertices in regular region
tends to be less destructive than removal in the non-regular region,
as also observed for the BA and NL network models in
Figure~\ref{Fig:cascade}.

While analyzing the effect of the regularity on the largest regular
regions in BA, NL and the three protein-protein interaction networks,
which are scale-free networks, we observed that there is a positive
correlation between the super-regularity coefficient and the average
of the relative size $G$ of the largest component for $0\leq \alpha
\leq 0.1$ --- the obtained Pearson coefficient is 0.7. This indicates
that the more regular the regions, the more robust they are under
removal of vertices in such regions. Moreover, we investigated the
effect of the variation of each four considered measurements
($k(i),cc(i),r(i)$ and $loc(i)$) and observed that each of them is
correlated to the average of the relative size $G$. Therefore, we can
conclude that the smaller the variation of the local measurements
inside the largest regular region, the more robust the network is with
respect to the cascade dynamics. This is a consequence of the
homogeneous distribution of shortest paths in such regions --- as the
vertices tend to have similar properties, their betweenness centrality
becomes similar. This effect was identified in every considered
networks. Indeed, vertices with the highest betweenness tend to be
outside the regular regions, since such vertices generally present
distinct local properties (such as the hubs).  These vertices tend to
be the outliers in the PCA projection.

\begin{table}[t]
\caption{The simplicity in different networks.}
\begin{tabular}{l|c|c|c|c|c|c}
\hline
{\bf Network} &    {\bf $N$} &    {\bf $\langle k \rangle$} &   {\bf $\langle cc \rangle$} &    {\bf $P$} & {\bf $\Sigma$} & $\Gamma$\\
\hline
\emph{S. cerevisiae}         &5439 &28.4 &0.20 &24 &0.13 & 0.44\\
\emph{D. melanogaster}   &7286 &6.84 &0.02 &38 &0.54 & 0.47\\
\emph{H. Sapiens}            &8792 &7.13 &0.07 &32 &0.75 & 0.60\\
\hline
Barab\'{a}si-Albert      &1000   &8  &0.04  & 34  & 0.45  &0.34  \\
Krapivsky-Redner         &1000   &8  &0.02  & 30 & 0.50  &0.36 \\
Erd\H{o}s-R\'{e}nyi     &1000   &8  &0.01  & 25  & 0.60  &0.43 \\
Knitted                           &1000   &8  &0.01  & 10  & 0.70  &1.0 \\
\hline
\end{tabular}
\label{table1}
\end{table}

The reported method for identifying regular subgraphs (or patches)
within complex networks, as well as the respectively obtained results,
illustrated and corroborated the importance of considering patchwise
regularity in order to characterize and obtain insights about the
properties of theoretical and real-world networks.  Several are the
possibilities for further investigations opened by this work, which
include but are not limited to the consideration of other real-world
networks, selection of additional topological measurements, and
analysis of evolution of simplicity in some biological networks, as
the brain, food webs and genetic networks. In addition, our
methodology can be applied to the identification of more general types
of motifs, defined by similar structures. These motifs may not have
fully regular structure, as defined currently~\cite{Milo:2002}, but be
characterized by statistically uniform properties.

\begin{acknowledgments}
Luciano da F. Costa thanks CNPq (308231/03-1) and FAPESP (05/00587-5) for sponsorship. Francisco Aparecido
Rodriges is grateful to FAPESP (07/50633-9).
\end{acknowledgments}

\bibliography{segnet-PRL}

\end{document}